\documentclass[3p,review,preprint,12pt,authoryear]{elsarticle}

\usepackage{amssymb}
\usepackage{gensymb}
\usepackage[utf8]{inputenc}
\usepackage{textcomp}
\usepackage{multirow}
\usepackage{amsmath}
\usepackage{geometry}
\usepackage{setspace}
\usepackage{array, booktabs, makecell}

\usepackage{siunitx}
\usepackage[version=4]{mhchem}
\usepackage{graphicx}

\journal{Icarus}

\begin{document}

\begin{frontmatter}


\title{Europa's surface water ice crystallinity: Discrepancy between observations and thermophysical and particle flux modeling}



\author[1]{Jodi R. Berdis\corref{cor1}}
\ead{berdis@nmsu.edu}

\author[2]{Murthy S. Gudipati} 
\author[1]{James R. Murphy}
\author[1]{Nancy J. Chanover}

\address[1]{Astronomy Department, New Mexico State University, Las Cruces, NM, USA}
\address[2]{Science Division, Jet Propulsion Laboratory, California Institute of Technology, Pasadena, CA, USA}

\cortext[cor1]{Corresponding author}


\begin{abstract}

Physical processing of Europan surface water ice by thermal relaxation, charged particle bombardment, and possible cryovolcanic activity can alter the percentage of the crystalline form of water ice compared to that of the amorphous form of water ice (the ``crystallinity'') on Europa’s surface. The timescales over which amorphous water ice is thermally transformed to crystalline water ice at Europan surface temperatures (80-130 K) suggests that the water ice there should be primarily in the crystalline form, however, surface bombardment by charged particles induced by Jupiter's magnetic field, and vapor deposition of water ice from Europan plumes, can produce amorphous water ice surface deposits on short timescales. The purpose of this investigation is to determine whether the Europan surface water ice crystallinity derived from ground-based spectroscopic measurements is in agreement with the crystallinity expected based upon temperature and radiation modeling. Using a 1D thermophysical model of Europa's surface, we calculate an integrated full-disk crystallinity of Europa's leading hemisphere by incorporating the thermal relaxation of amorphous to crystalline water ice and the degradation of crystalline to amorphous water ice by irradiation. Concurrently, we derive the full-disk crystallinity of Europa's leading hemisphere using a comparison of near-infrared ground-based spectral observations from \citet{grundy+1999}, \citet{busarev+2018}, and the Apache Point Observatory in Sunspot, NM, with laboratory spectra from \citet{mastrapa+2008} and the Ice Spectroscopy Lab at the Jet Propulsion Laboratory. We calculate a modeled crystallinity significantly higher than crystallinities derived from ground-based observations and laboratory spectra. This discrepancy may be a result of geophysical processes, such as by vapor-deposited plume material, or it may arise from assumptions and uncertainties in the crystallinity calculations.

\end{abstract}


\end{frontmatter}


\section{Introduction}
\label{sec:intro}

Water ice exists in both the crystalline (hexagonal or cubic) and amorphous phases in conditions prevalent in the Solar System \citep[e.g.,][]{hobbs1974, jenniskens+1998, petrenko+whitworth1999}. For small, inactive Solar System objects with surface temperatures greater than $\sim75$ K, surface water ice is expected to currently be in the crystalline phase even if it existed in the amorphous phase at the time of the object's formation \citep{kouchi+1994}. The amorphous form of water ice is a metastable phase and is most commonly detected in the interstellar medium, where it is produced by direct condensation of water vapor at temperatures as low as 10 K \citep[e.g.,][and references therein]{jenniskens+1995}. Amorphous water ice relaxes into a crystalline structure at a temperature-dependent rate, where the relaxation is much faster at warmer temperatures. For example, amorphous water ice at T $\sim90$ K will relax into a crystalline structure after $10^{5}$ years, whereas amorphous water ice at T $\sim115$ K will relax into a crystalline structure after a few years \citep{kouchi+1994, jenniskens+1998, mastrapa+2013}.

A surface that experiences additional ice transformation processes, such as thermal evaporation, ion sputtering, radiation-induced sublimation, plumes, etc., may experience an ice phase change in its existing surface and in redeposited frost from the gas phase. These processes may result in spatially and temporally varying amounts of crystalline water ice compared to the combined content of amorphous and crystalline water ice (hereafter referred to as the ``crystallinity percentage,'' or simply, ``crystallinity''). Charged particle bombardment from a gas giant's magnetic field, for example, can cause disruption and disorder of the surface ice's crystalline structure, producing an amorphous-like structure at a rate dependent on the particle bombardment flux \citep{cooper+2001, baragiola+2003}. Additionally, tidal flexing may produce cracks in the surface ice that could vent out flash-frozen plume material, coating localized regions of the surface with vapor-deposited amorphous water ice \citep{kouchi+1994}, thereby decreasing the region's crystallinity. While similarities in the near-infrared spectra of radiation-altered crystalline and vapor-deposited amorphous water ice is the primary factor behind interpreting this phenomenon as ``amorphization of crystalline ice,'' to date there exists no direct evidence indicating that the crystalline structure is completely destroyed during this process. Furthermore, radiation-altered ice and vapor-deposited amorphous ice may not have similar bulk density and porosity. Therefore, we should use caution when referring to the radiation-processed crystalline water ice as amorphous water ice, and rather should use the term amorphous-like ice. Refrosting from gas-phase water onto a cold surface at temperatures well below 120 K will also result in emplacement of amorphous ice, so these processes should be distinguished in interpreting the spectral signatures of icy bodies.

Europa is likely affected by the aforementioned ice transformation processes. Its subsurface liquid water ocean \citep{carr+1998, pappalardo+1999} influences the surface directly due to upwelling diapirs, plumes, other forms of cryovolcanic activity \citep{pappalardo+1998, collins+nimmo2009, roth+2014}, and tectonic resurfacing \citep{greenberg+1998, pappalardo+1999}. The age of Europa's surface is therefore often ``reset,'' where the current age estimate of its surface is $\sim10^{7}$--$10^{8}$ years old \citep{zahnle+1998, pappalardo+1999, bierhaus+2009}. The crystallinity percentage of Europa's surface is a balance between: (1) thermal relaxation of amorphous to crystalline water ice; (2) conversion of crystalline to amorphous water ice due to charged particle bombardment from Jupiter's magnetic field; (3) possible vapor deposition of amorphous water ice as plume material; and (4) any additional ongoing cryovolcanic activity at the surface, such as diapirs or other processes that may form chaotic terrain \citep{collins+nimmo2009}, that could partially warm or melt the surface ice and convert amorphous to crystalline water ice. Thus, investigating the crystallinity of Europa's surface water ice can provide insight into the active processes that affect its surface.

Spectroscopic observations are a commonly used tool for assessing the crystallinity percentage of a surface. Water ice absorption features in the near-infrared, such as those at $1.5$ and $1.65$ $\mu$m, increase in depth and shift to longer wavelengths as the crystalline fraction increases \citep{schmitt+1998, mastrapa+2008}. Optical constants and laboratory spectra of pure crystalline and amorphous water ice at temperatures between $18-270$ K \citep[e.g.,][]{grundy+schmitt1998, mastrapa+2008, mastrapa+2009} provide reference spectra for comparison to, and interpretation of, observed spectra of Solar System objects such as Europa. \citet{hansen+mccord2004} and \citet{ligier+2016} sought to constrain approximate fractions of crystalline and amorphous water ice on Europa's surface. \citet{hansen+mccord2004} suggested that at a depth of $\sim1$ mm (i.e., the depth probed by near-infrared observations) and deeper, water ice on Europa is predominantly crystalline, however nearly all of the water ice above a depth of $\sim1$ mm is amorphous. \citet{ligier+2016} found that crystalline water ice makes up on average $40-55\%$ of all material (including non-ice material) on the leading hemisphere, where the amorphous-to-crystalline water ice ratio is 0.57 globally, indicating a crystallinity of $64\%$. The fraction of water ice relative to non-ice material on Europa's leading hemisphere is greater than that found on the trailing hemisphere; \citet{dalton+2012} find a high abundance ($\sim70-80\%$) of total water ice compared to non-ice materials in several Galileo Near-Infrared Mapping Spectrometer observations of the leading hemisphere, whereas the trailing hemisphere likely only has $\sim20-30\%$ \citep{dalton+2012}. In order to minimize the spectral effects from brines or other non-ice materials and maximize the water ice signal, we focus solely on Europa's leading hemisphere in this study.

Near-infrared (NIR, $\sim1.25-2.5 \;\mu$m) ground-based observations of Europa have revealed bulk properties on a hemispheric scale, such as temperature \citep[e.g.,][]{grundy+1999} and composition \citep[e.g.,][]{busarev+2018}. Spatially-resolved spectral compositional maps of various brines and radiation products have also been produced from observations in this wavelength regime and have revealed compositional dependencies of surface features and hemispheres \citep[e.g.,][]{spencer+2006, brown+hand2013, hand+brown2013, ligier+2016}. These studies focused largely on non-ice species on the surface rather than the water ice components. However, \citet{ligier+2016} also produced crystalline and amorphous water ice maps and suggested that the relative fraction of crystalline and amorphous water ice on the surface can elucidate the balance between crystallization due to thermal relaxation and amorphization due to particle bombardment. We investigated this balance in our study using full-disk (rather than spatially resolved) data in order to reduce complications due to compositional variations across the surface and focus on the bulk properties of the leading hemisphere.

Numerical modeling can provide an evaluation of the evolution of icy planetary surfaces under the influence of a variety of physical processes such as thermal relaxation and charged particle flux. One-dimensional thermophysical modeling allows the simulation of surface temperature cycles based on the orbital characteristics and surface properties of an object, such as albedo, thermal inertia, emissivity, etc. \citep[e.g.,][]{spencer+1999,rathbun+2010, trumbo+2017, trumbo+2018}. \citet{baragiola+2013} modified a methodology from \citet{fama+2010} for determining the fraction of an assumed $100\%$ crystalline water ice surface that has been converted into amorphized ice as a function of radiation dose, elapsed time, and surface temperature. \citet{dalleore+2015} determined the approximate ages of two craters on Saturn's moon Rhea by calculating the crystallinity percentage of the craters based on the depth and position of the $2.0\;\mu$m band, then utilizing the amorphous fraction methodology from \citet{baragiola+2013}. In a similar manner, we compute the fraction of amorphized ice on Europa's leading hemisphere given the radiation dose, temperature, and age of Europa's surface, and excluding transformative geophysical processes such as cryovolcanism.

In \S\ref{sec:obs}, we derive the observed crystallinity of Europa's leading hemisphere using NIR ground-based observations from \citet{grundy+1999}, \citet{busarev+2018}, and the Apache Point Observatory (APO) in Sunspot, NM, and laboratory spectra from \citet{mastrapa+2008} and the Ice Spectroscopy Lab (ISL) at the Jet Propulsion Laboratory (JPL) in Pasadena, CA. In \S\ref{sec:model}, using a 1D thermophysical model \citep{spencer+1989} of Europa's surface, we calculate the expected full-disk crystallinity of Europa's leading hemisphere by incorporating the thermal relaxation of amorphous to crystalline water ice and the formation of amorphous water ice by irradiation. In \S\ref{sec:res_disc}, we assess the results from the observed and modeled crystallinity calculations and discuss possible sources for their discrepancies. Finally, in \S\ref{sec:concl}, we provide concluding remarks about our findings.

\section{Observed Crystallinity}
\label{sec:obs}

In order to calculate the water ice crystallinity of Europa's leading hemisphere, we employ both ground-based observations of Europa's illuminated leading hemisphere and also laboratory data of crystalline and amorphous water ice at a temperature that corresponds to the average surface temperature of Europa's leading hemisphere at the time of observation. This ensures that temperature dependencies of the $1.5$ and $1.65$ $\mu$m bands \citep[e.g.,][]{grundy+schmitt1998, mastrapa+2008} are properly accounted for in our calculations of the crystallinity. We describe the ground-based observations in \S\ref{sec:ground-data}, the laboratory experiments in \S\ref{sec:lab-data}, and our crystallinity calculation methods in \S\ref{sec:methods}.

\subsection{Ground-Based Data Description}
\label{sec:ground-data}

Both \citet{grundy+1999} and \citet{busarev+2018} acquired ground-based, full-disk, spectroscopic observations of the leading and trailing hemispheres of several icy outer Solar System satellites, including Europa. \citet{grundy+1999} obtained surface ice temperatures using the strengths of notable water ice absorption bands in the $1.5-1.65$ $\mu$m region, whereas \citet{busarev+2018} investigated spectral and compositional differences between the leading and trailing hemispheres. In order to ensure that no drastic events have occured to significantly change the crystallinity by a detectable amount over time from 1998 \citep{grundy+1999} to 2016/2017 \citep{busarev+2018} to 2018 (this work), we obtain additional NIR reflectance spectra of Europa's leading hemisphere and supplement our observations with those of \citet{grundy+1999} and \citet{busarev+2018} in the crystallinity calculation outlined in \S\ref{sec:methods}.

\begin{figure}[!ht]
 \centering
 \includegraphics[scale=0.5]{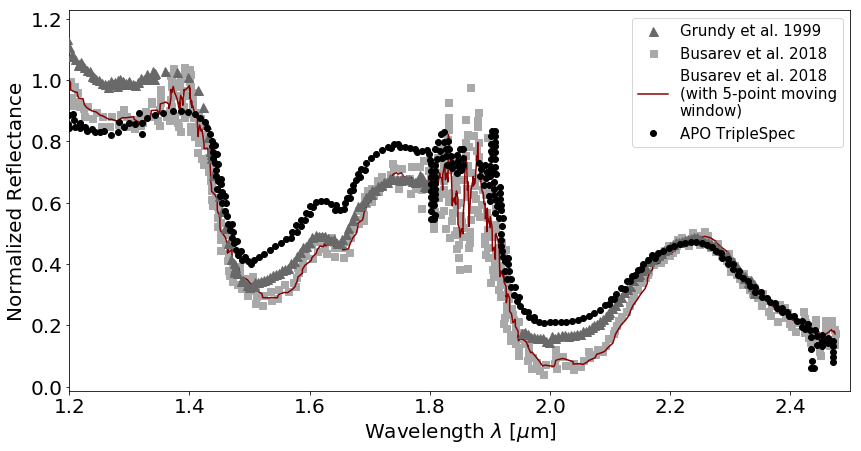}
 \caption{Ground-based observations from \cite{grundy+1999} (dark gray triangles), \cite{busarev+2018} (light gray squares), \cite{busarev+2018} with the 5-point moving window (dark red solid line), and APO TripleSpec (black circles). Atmospheric water vapor absorption in the $1.35-1.43\;\mu$m and $1.80-1.92\;\mu$m regions led to either data gaps or extremely high noise and should be flagged as untrustworthy regions. Uncertainties that are not in those regions are smaller than the size of the data points.}
 \label{fig:fullspec}
\end{figure}

We obtained NIR spectra of Europa with the Astrophysical Research Consortium 3.5 m telescope at Apache Point Observatory (APO) using TripleSpec, a cross-dispersed near-infrared spectrograph, with a spectral resolution of $R= \lambda/\Delta\lambda= 3500$ in the wavelength range $\lambda = 0.96-2.47 \;\mu$m \citep{wilson+2004}. The telescope was nodded to place Europa on two positions within the $1.1''$ $\times$ $43''$ slit in order to improve sky subtraction. We observed Europa's leading hemisphere on 11 May 2018, when it had a phase angle of $0.67\degree$ and an angular diameter of $0.98''$. We reduced and combined 126 consecutive spectra using Triplespectool, a modified version of the Spextool data reduction package \citep{vacca+2003,cushing+2004}. We used the G2V star HR 4328 as our standard star to remove telluric contributions and the solar spectrum. Significant atmospheric water vapor absorption exists in the $1.35-1.43\;\mu$m and $1.80-1.92\;\mu$m regions. While atmospheric water vapor absorption at $\sim1.9 \;\mu$m does not impact our study, a portion of the $\sim1.4 \;\mu$m region is used in the band integration described in \S\ref{sec:methods}. We therefore fit a continuum to this region ($\sim1.35-1.43 \;\mu$m) for the \citet{grundy+1999} and APO TripleSpec spectra and used the continuum-fitted data in lieu of the original data impacted by atmospheric water vapor absorption. The spectrum from \citet{busarev+2018} included more noise overall than the \citet{grundy+1999} and APO TripleSpec spectra, so we apply a 5-point moving window and include the smoothed spectrum along with the original \citet{busarev+2018} spectrum in our crystallinity analysis below. Our reduced leading hemisphere spectrum acquired from APO's TripleSpec instrument is presented, along with full-disk leading hemisphere spectra from \cite{grundy+1999} and \cite{busarev+2018}, and the 5-point moving window spectrum from \cite{busarev+2018}, in Figure \ref{fig:fullspec}. Slight differences in slopes and absorption band depths are likely a result of different normalization points and/or polynomial continuum removal.

\subsection{Laboratory Data Description}
\label{sec:lab-data}

\begin{figure}[!ht]
 \centering
 \includegraphics[scale=0.5]{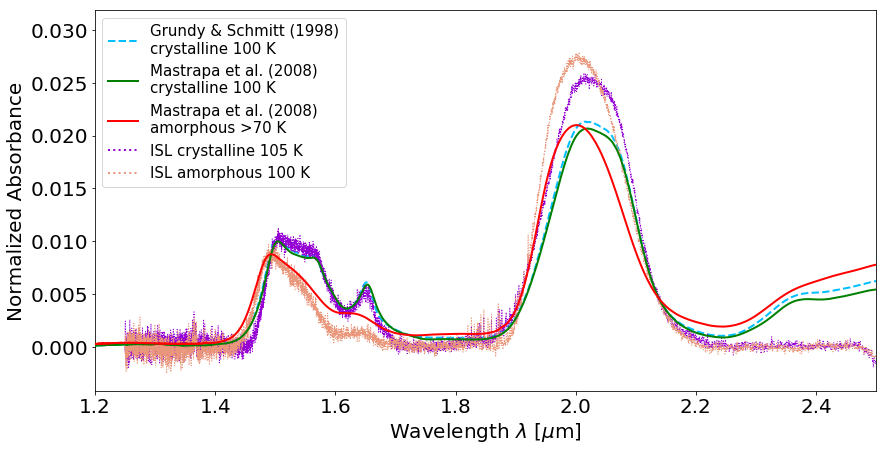}
 \caption{Normalized absorbance spectra of pure amorphous and crystalline water ice from \citet{mastrapa+2008} (red and green solid lines) and the ISL (orange and purple dotted lines), and pure crystalline water ice from \citet{grundy+schmitt1998} (blue dashed line).}
 \label{fig:labspec}
\end{figure}

We produce absorbance spectra of crystalline and amorphous water ice from the $n$ and $k$ extinction coefficients published in \citet{mastrapa+2008} and the $\alpha$ absorption coefficient published in \citet{grundy+schmitt1998} as an additional reference. Since optical constants of both crystalline and amorphous water ice at Europa temperatures in the NIR have only been published by a couple of groups in the past $\sim20$ years \citep{mastrapa+2008, grundy+schmitt1998}, we include an additional transmission dataset of crystalline and amorphous water ice from the ISL at JPL. The ISL reproduces surface environmental conditions (e.g., particle radiation bombardment, temperature, and composition) of the surfaces of various Solar System objects, such as comets and Europa \citep[e.g.,][]{barnett+2012,lignell+gudipati2015,gudipati+2017DPS}. NIR spectra ($1.25 - 4\;\mu$m) of pure water ice were acquired in $2010$ by both depositing water ice from vapor at constant temperatures ranging from $18$ to $140$ K, and by ceasing deposition while varying the temperature. We choose crystalline and amorphous spectra obtained under a common ice thickness and temperatures close to $100$ K, which we assume to be the approximate average surface temperature of Europa at the time of our ground-based observations. Absorbance is calculated as $log(T_{0}) - log(T)$, where $T$ is the sample transmission and $T_{0}$ is the background transmission; polynomial continua are removed from both crystalline and amorphous water ice spectra. We then normalize all three sets of absorbance data (from \citet{mastrapa+2008}, \citet{grundy+schmitt1998}, and the ISL), as shown in Figure \ref{fig:labspec}.

\subsection{Integrated Band Area Ratios}
\label{sec:methods}

Several absorption bands in the NIR serve as useful diagnostics for calculating the crystallinity percentage of an ice sample or surface. The water ice absorption features at $1.5$ and $1.65$ $\mu$m increase in depth and shift to longer wavelengths as the crystalline fraction increases \citep{schmitt+1998,mastrapa+2008}. Therefore, observed spectra of Solar System objects such as Europa may be compared to those of laboratory-produced pure crystalline and amorphous water ice to approximate the crystallinity percentage. However, crystalline water ice spectra at high temperatures in the NIR region look very similar to amorphous water ice spectra at lower temperatures \citep{grundy+schmitt1998}, so variations in local surface temperatures may be incorrectly interpreted as being due to amorphous water ice.

Calculating the $1.65/1.5$ $\mu$m integrated band ratios of laboratory-produced pure crystalline and amorphous water ice at a temperature similar to that of Europa's surface provides a reference against which Europan surface crystallinity may be derived from ground-based and future space-borne (such as the MISE instrument aboard Europa Clipper) spectroscopic observations. \citet{mastrapa+2008} find the $1.65/1.5$ $\mu$m integrated band area ratio to be one of the best metrics for distinguishing between crystalline and amorphous water ice. The $1.65$ $\mu$m band has often been used as an estimator for the approximate fraction of crystalline water ice on icy surfaces \citep[e.g.,][etc.]{grundy+1999, jewitt+luu2004, cook+2007}, assuming relatively high confidence in the ice temperature. Additionally, water ice absorption bands in the NIR increase in depth with increasing grain size \citep{clark1981}. We use a ratio of the $1.65$ $\mu$m and $1.5$ $\mu$m integrated band areas, rather than just the area or depth of the $1.65$ $\mu$m band, in order to remove the effects of grain size dependency in our crystallinity calculations. We therefore are not limited by grain size assumptions, unless the surface is dominated by significantly large grains (e.g., $>500\;\mu$m, \citealp{clark1981}), at which point the spectra cannot be further quantified due to band saturation.

\begin{figure}[!ht]
 \centering
 \includegraphics[scale=0.5]{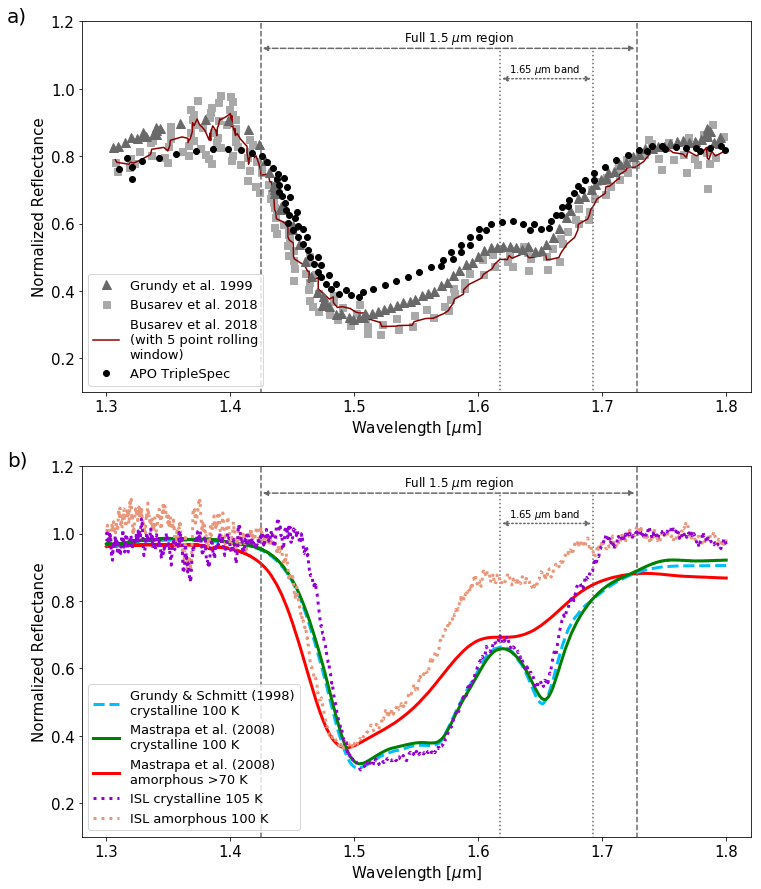}
 \caption{(a) Full-disk ground-based spectra in the $1.5$ $\mu$m region of Europa’s leading hemisphere from \citet{grundy+1999} (dark gray triangles), \citet{busarev+2018} (light gray squares), \cite{busarev+2018} with the 5-point moving window (dark red solid line), and APO TripleSpec (black circles). Colors and point styles  match those in Figure \ref{fig:fullspec}. (b) Laboratory spectra of pure amorphous and crystalline water ice from the ISL at JPL (orange and purple dotted lines) and \citet{mastrapa+2008} (red and green solid lines), and pure crystalline water ice from \citet{grundy+schmitt1998} (blue dashed lines). Dark gray vertical dashed lines indicate the bounds of integration for the $1.5$ $\mu$m region, and light gray vertical dotted lines indicate the bounds of integration for the $1.65$ $\mu$m region. Colors and point styles  match those in Figure \ref{fig:labspec}.}
 \label{fig:1.5region}
\end{figure}

The boundaries for our integration of band depths for both the \citet{mastrapa+2008} and ISL spectra were chosen to be the same as used by \citet{mastrapa+2008} for consistency ($1.424-1.731$ $\mu$m for the $1.5$ $\mu$m region, and $1.618-1.695$ $\mu$m for the $1.65$ $\mu$m band at $100$ K; see Figure \ref{fig:1.5region}). We fit straight lines to 2-3 points at each edge of the bounds and integrated the absorption profile under the straight line fit. The integrated band area ratio ($B$) is calculated as the integrated area of the $1.65$ $\mu$m band divided by the integrated area of the $1.5$ $\mu$m region, where a larger $B$ corresponds to a higher crystallinity.

The $1.65/1.5$ $\mu$m integrated band area ratio of the ground-based observations is a linear combination \citep[similar to that used by][]{hansen+mccord2004} of the crystalline and amorphous laboratory integrated band area ratios, or:
\begin{equation}
 B_{GBO} = C \times B_{cryst} + (1 - C) \times B_{amorph},
 \label{eq:bar}
\end{equation}

\noindent where $B_{GBO}$ is the $1.65/1.5$ $\mu$m integrated band area ratio using the ground-based observations (GBO), $B_{cryst}$ is the $1.65/1.5$ $\mu$m integrated band area ratio using the laboratory-acquired crystalline end-member spectrum, $B_{amorph}$ is the $1.65/1.5$ $\mu$m integrated band area ratio using the laboratory-acquired amorphous laboratory spectrum, and $C$ is the crystallinity of the ground-based observations for which we are solving. We computed the crystallinity percentages of the three full-disk ground-based observations from both the \citet{mastrapa+2008} and the ISL absorbance spectra (Table \ref{table:obs_cryst}). All values have $\sim10\%$ errors due to the uncertainties in the laboratory ice sample thicknesses \citep{mastrapa+2008}. Previous estimates of Europa's leading hemisphere crystallinity fall in the range of $~40-60\%$ \citep{hansen+mccord2004,ligier+2016}. Crystallinities derived from the \citet{grundy+1999}, \citet{busarev+2018} 5-point moving window, and APO TripleSpec ground-based observations are slightly smaller than these previous estimates, whereas crystallinities derived from the original \citet{busarev+2018} data are slightly larger than previous estimates. Such a large increase in crystallinity from 1999 to 2016/2017 and an equally large decrease from 2016/2017 to 2018 is unlikely due to any physical surface processes or changes in the thermal and/or ion environment. The original spectrum of \citet{busarev+2018} has more noise than the spectra from \citet{grundy+1999} and APO TripleSpec, which gives inaccurate results for the integrated band area ratios. The 5-point moving window provides better agreement with the \citet{grundy+1999} and APO TripleSpec observations.

\begin{spacing}{1.0}
\begin{table}[ht]
\caption{Full-disk crystallinities for the three ground-based observations each derived using the \citet{mastrapa+2008} and the Ice Spectroscopy Lab crystalline and amorphous water ice spectra. }
\centering
\scriptsize
  \begin{tabular}{c|c c c c}
  \hline
   & \thead{\citet{grundy+1999} \\ B$_{GBO}=0.0336$} & \thead{\citet{busarev+2018} \\ B$_{GBO}=0.0548$} & \thead{\citet{busarev+2018} \\ (5-point window) \\ B$_{GBO}=0.0353$} & \thead{TripleSpec \\ B$_{GBO}=0.0350$} \\ \hline
  \thead{\citet{mastrapa+2008} \\ B$_{cryst}=0.0741$ \\ B$_{amorph}=0.0182$}	& $27.5\%$ & $65.5\%$ & $30.5\%$ & $30.1\%$  \\ \hline
  \thead{Ice Spectroscopy Lab \\ B$_{cryst}=0.0715$ \\ B$_{amorph}=0.0144$}	& $33.6\%$ & $70.8\%$ & $36.5\%$& $36.1\%$  \\ \hline
  
  \end{tabular}
\label{table:obs_cryst}
\end{table}
\end{spacing}


\section{Modeled Crystallinity}
\label{sec:model}

\citet{baragiola+2013} modified a methodology from \citet{fama+2010} for determining the fraction of an assumed $100\%$ crystalline water ice surface that has been converted into amorphized ice due to a radiation dose over a given timescale at a given temperature. One-dimensional thermophysical modeling allows the simulation of daily surface temperature cycles based on the orbital characteristics and surface properties of an object, such as albedo, thermal inertia, emissivity, etc. \citep[e.g.,][]{spencer+1999,trumbo+2017,trumbo+2018}. This allows for the determination of simulated surface temperatures over a complete orbit, and therefore, a first-order approximation of the spatially-dependent average surface temperatures over long timescales. The rate of thermal relaxation of amorphous to crystalline water ice can be inferred based on the present-day surface temperatures of an object \citep{kouchi+1994,jenniskens+1998}, assuming, in the case of Europa, that the present-day surface temperatures are also representative of temperatures throughout the past $10^{7}$--$10^{8}$ years. The expected crystallinity percentage may then be calculated based on the fraction of radiation-induced amorphized ice, provided the temperature of the surface is constrained to within a few K, since significant changes in the NIR spectra are noticed at higher temperatures of crystalline ice that may resemble amorphous ice features at lower temperatures.

While Europa's surface water ice crystallinity as derived from ground-based spectra will have a dependence on contributions from thermal relaxation, particle radiation, and geophysical processes, the crystallinity as calculated from this numerical modeling technique will only include contributions from thermal relaxation and particle radiation, since additional contributions from geophysical processes are much less constrained. If discrepancies exist between the crystallinities derived from numerical modeling and ground-based observations, the disparity may be a result of geophysical processes, such as by vapor-deposited plume material or sub-surface diapirs, or assumptions and uncertainties in the crystallinity calculations.

For this investigation we developed the Incipient Code for Investigating the Crystallinity of the Leading-hemisphere of Europa (ICICLE). ICICLE employs an adaptation of a 1D numerical thermophysical model for airless bodies \citep{spencer+1989} that has been adjusted to reflect orbital characteristics of Europa and environmental and surface conditions of its leading hemisphere. Spatially varying thermal inertia, emissivity, and albedo values employed are those presented in Figure 3 of \citet{trumbo+2018}. Their albedo values at a $0.5\degree$ latitude-longitude spatial resolution were extrapolated from albedo measurements made by \emph{Voyager}, and their thermal inertia and emissivity values were derived via modeling at a $3\degree$ latitude-longitude spatial resolution based on four $233$ GHz ($1.3$ mm) observations acquired with the Atacama Large Millimeter Array (ALMA). These thermal inertia and emissivity data were modeled only for locations within $57\degree$ of the central point of each ALMA observation, where regions beyond that were not modeled. We include uncertainties in thermal inertia, emissivity, and albedo of $\pm 5\%$, $\pm 5\%$, and $\pm 0.03$, respectively. We produce latitudinal averages at each longitude to latitudinally extend the thermal inertia and emissivity data beyond those presented in \citet{trumbo+2018} in order to obtain approximate values for those higher latitude locations for which there were no observations. We also resampled the albedo data from a $0.5\degree$ spatial resolution to a $3\degree$ spatial resolution in order to match the spatial sampling of the thermal inertia and emissivity data.

\begin{figure}[!ht]
 \centering
 \includegraphics[scale=0.6]{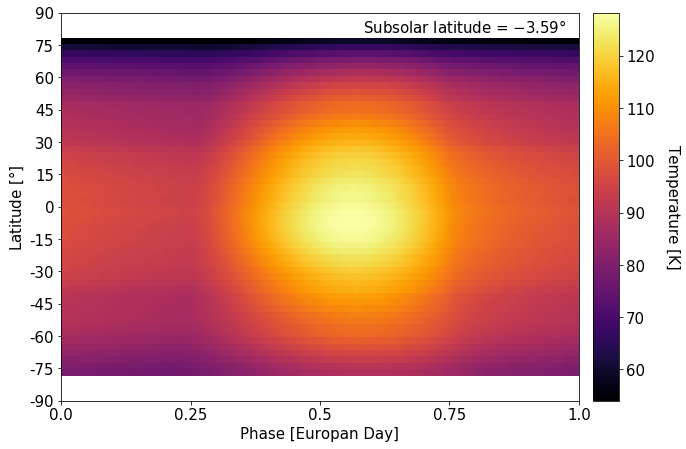}
 \caption{ICICLE-simulated surface temperatures as a function of latitude and phase throughout one Europan day at a heliocentric distance of 5.408 AU and a subsolar latitude of $-3.59\degree$, corresponding to the orbital parameters of Europa on 11 May 2018 during our ground-based observations.}
 \label{fig:temp_latvstod}
\end{figure}

Jupiter has an axial tilt of $\sim3\degree$ with respect to the solar ecliptic, and our model predicts that, at a particular time of day and location on Europa's surface, the surface temperature changes by $\sim3-6$ K throughout the year as a result of this tilt. We conducted four thermophysical simulations of Europa corresponding to four positions throughout Jupiter's orbit around the Sun in order to include seasonal variations into our temperature predictions. The orbital parameters of the first position (Orbital Position $\#1$ in Table \ref{table:var_params}) were chosen to coincide with the ground-based observations on 11 May 2018. The remaining three orbital positions were conducted one quarter of Jupiter's orbit in time after each previous position (Table \ref{table:var_params}). After each thermophysical model run had thermally equilibrated to energy conservation (i.e., the total solar insolation was equal to the outward emitting power), we generated simulated surface temperature cycles over one Europan rotation period (Figure \ref{fig:temp_latvstod}) at the four orbit locations, which were then averaged to simulate a full Europan year. We assumed a constant regolith density, $\rho_{reg}$, of $0.5$ g cm$^{-3}$ \citep{spencer+1999}, a specific heat of surface water ice, $c_{p}$, of $9 \times 10^{6}$ erg g$^{-1}$ K$^{-1}$ \citep{feistel+wagner2006,trumbo+2017}, and a model time step of $1/2500$ of a Europa day ($2.55$ Earth days).

\begin{spacing}{1.0}
\begin{table}[ht]
\caption{Orbital parameters for the four thermophysical simulations to account for thermal variation throughout a Europan year.}
\centering
\footnotesize
  \begin{tabular}{c c c}
  \hline
  \thead{Orbital \\ Position \#} & \thead{Heliocentric \\ Distance [AU]} & \thead{Subsolar \\ Latitude [\textdegree]} \\ \hline
  $1$	& $5.408$			& $-3.589$  \\
  $2$	& $5.005$			& $+3.770$  \\
  $3$	& $5.057$			& $-0.466$  \\
  $4$	& $5.364$			& $-0.195$  \\ \hline
  
  \end{tabular}
\label{table:var_params}
\end{table}
\end{spacing}

Using the derived average surface temperatures at each $3\degree \times 3\degree$ latitude-longitude model grid location on Europa's surface, we used ICICLE to compute the fraction of water ice that was converted from the crystalline to the amorphous phase due to a particle radiation dose over a given timescale. This method, first introduced by \citet{fama+2010} and adapted by \citet{baragiola+2013} and \citet{dalleore+2015}, is given in equation form as

\begin{equation}
 \varPhi_{A} = \varPhi_{A_{max}} \Big( 1 - \textrm{exp}\Big[\frac{-kFt}{N}\Big] \Big).
 \label{eq:amorph}
\end{equation}

\begin{figure}[!ht]
 \centering
 \includegraphics[scale=0.6]{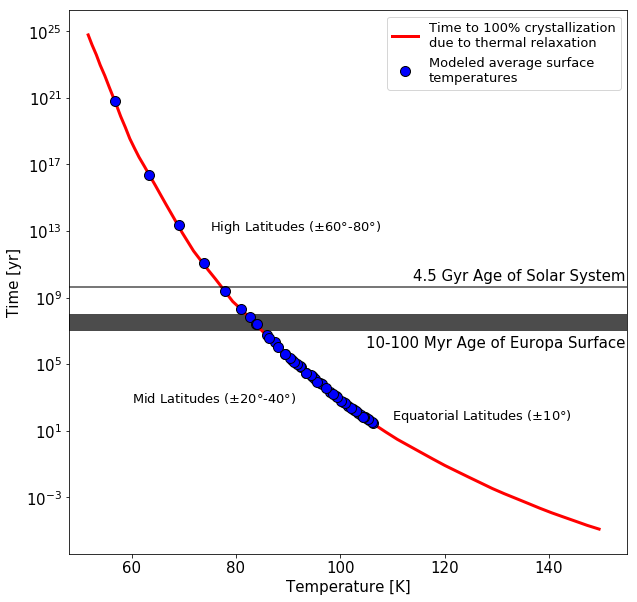}
 \caption{Time to 100\% crystallization of water ice due to thermal relaxation. The red solid line represents the amount of time required to convert amorphous water ice into $100\%$ crystalline water ice as a function of temperature \citep{kouchi+1994, mastrapa+2013}, and the blue circles are our modeled average surface temperatures at each latitude band over one full jovian orbit. Each blue circle represents a longitudinal average at each latitude point to show how the temperature and time to $100\%$ crystallization change with latitude.}
 \label{fig:t_cryst}
\end{figure}

\noindent In Equation \ref{eq:amorph}, $\varPhi_{A}$ is the amorphous fraction for which we are solving, $\varPhi_{A_{max}}$ is the maximum fraction of amorphous water ice on the surface ($0.9-1.0$, following methods in \citealp{dalleore+2015}), and $k$ is a fitting parameter dependent on water ice temperature \citep[Figure 16.3 in][]{baragiola+2013,fama+2010}. To determine the particle radiation flux, $F$, we calculate the total number of protons, oxygen ions, and sulfur ions that are imparted upon the surface. While electrons impart a higher overall flux onto Europa's surface, protons and heavy ions have a higher propensity to sputter the surface at the few $\mu$m - mm depths. Since proton- and ion-induced amorphization is therefore far more important in this scenario, we only include proton flux and sulfur and oxygen ion fluxes. Furthermore, we consider any heating of the surface ice caused by ion or electron bombardment to be negligible. \citet{cooper+2001} compute particle number fluxes imparted onto Europa's incidence surface as $1.5 \times 10^{7}$ protons cm$^{-2}$ s$^{-1}$, $1.5 \times 10^{6}$ oxygen ions cm$^{-2}$ s$^{-1}$, and $9.0 \times 10^{6}$ sulfur ions cm$^{-2}$ s$^{-1}$ \citep[Table 2 of][]{cooper+2001}. We sum these three particle fluxes and estimate a total particle flux of $2.55 \times 10^{7}$ particles cm$^{-2}$ s$^{-1}$. This should be considered an upper limit since these particle number fluxes reflect the ion environment at Europa's approximate distance from Jupiter and ion fluxes decrease from the trailing hemisphere to the leading hemisphere \citep{pospieszalska+johnson1989}. In Equation \ref{eq:amorph}, $N$ is the number of H$_{2}$O molecules contained within a surface volume that is $1$ cm$^{2}$ in cross-section and $\sim5-30$ $\mu$m in vertical extent for which hydrogen, oxygen, and sulfur ions may bombard the surface, where $5$ $\mu$m is the approximate penetration depth for $1$ MeV oxygen and sulfur ions, and $30$ $\mu$m is the approximate penetration depth for $1$ MeV protons \citep[Figure 12(a) in][]{cheng+1986}. The radiation exposure time, $t$, represents the amount of time that initially pure crystalline water ice is exposed to radiation. For surface domains in the warmer equatorial regions, where surface temperatures exceed $\sim 120$ K, the time to $100\%$ crystallization due to thermal relaxation can be quite short, i.e., on the order of several days, and can be even shorter for surfaces with higher porosity \citep{mitchell+2017}. The radiation exposure time ($t$) is therefore also only several days since this is the timescale over which the surface ``resets'' to pure crystalline water ice. In cooler mid-latitude regions, thermal relaxation to 100\% crystallinity may take $10^{5}$ years, likewise setting the radiation exposure time, $t$, to $10^{5}$ years. In cold high-latitude regions, the time to thermal relaxation may be older than the age of Europa's surface ($10^{7}$--$10^{8}$ years), where the surface water ice age has been ``reset'' due to geophysical, tectonic, or other processes instead of thermal relaxation. The age of the surface (and therefore, the radiation exposure time) is set to $10^{7}$--$10^{8}$ years for model gridpoints with average temperatures that correspond to a thermal relaxation timescale larger than the estimated age of Europa's surface ($10^{7}$--$10^{8}$ years). Figure \ref{fig:t_cryst} shows the time to $100\%$ crystallization due to thermal relaxation for our modeled daily average surface temperatures.

\begin{spacing}{1.0}
\begin{table}[!ht]
\caption{ICICLE-derived crystallinities for several cases of differing thermal inertias (TI) and emissivities (e) in the latitudinally-averaged, extended filled regions, and differing temperatures (T) and fluxes (F) to explore the full range of the parameter space.}
\centering
\small
  \begin{tabular}{c l l}
  \hline
   Case & Method & Crystallinity \\ \hline
   1 & Filled TI/e 				& $89.8 \pm 8.3\;\%$  \\
   2 & Filled TI/e with TI$+10\%$, e$-10\%$ 	& $90.6 \pm 8.1\;\%$  \\
   3 & Filled TI/e with TI$-10\%$, e$+10\%$ 	& $89.5 \pm 8.4\;\%$  \\
   4 & Filled TI/e with max TI, min e 		& $90.8 \pm 8.1\;\%$  \\
   5 & Filled TI/e with min TI, max e 		& $87.2 \pm 7.1\;\%$  \\
   6 & Filled TI/e with T=T$+3$ K		& $94.4 \pm 4.8\;\%$  \\
   7 & Filled TI/e with T=T$-3$ K		& $83.0 \pm 13.3\;\%$  \\
   8 & Filled TI/e with F=F$*10$		& $81.1 \pm 14.6\;\%$  \\
   9 & Filled TI/e with F=F$/10$		& $95.0 \pm 4.4\;\%$  \\ \hline

  \end{tabular}
\label{table:model_cryst}
\end{table}
\end{spacing}

Crystallinity fractions at each model gridpoint were calculated as $1 - \varPhi_{A}$. We computed a projection area-weighted average of the crystallinity at each model gridpoint to derive a full-disk crystallinity (case 1) of the leading hemisphere as viewed from Earth to compare with the full-disk crystallinities derived from ground-based spectroscopic observations and laboratory spectra, as discussed in \S\ref{sec:obs}. For extended thermal inertia and emissivity data that were computed from latitudinal averages of the data derived in \citet{trumbo+2018}, we ran additional models to account for any errors that resulted from this averaging technique by altering the latitudinally-averaged thermal inertia and emissivity by $\pm 10\%$ (cases 2 and 3). We also replaced all latitudinally-averaged thermal inertia and emissivity values by their minimum and maximum values to test extreme conditions of the parameter space (cases 4 and 5). To account for uncertainties in temperature and flux, we ran several models whereby we increased and decreased all temperature values at each grid point by $3$ K (cases 6 and 7), and increased and decreased the particle radiation flux by a factor of 10 (cases 8 and 9). The derived crystallinities for all of the above mentioned cases are provided in Table \ref{table:model_cryst}. When the thermal inertia and emissivity are varied, our calculated range of crystallinities are $\sim87$--$91\%$; when temperature is varied, our calculated range of crystallinities are $\sim83$--$95\%$; and when particle radiation flux is varied, our calculated range of crystallinities are $\sim81$--$95\%$. Changes in these crystallinity of a few percent indicate that the ICICLE-modeled crystallinity is not substantially sensitive to the uncertainty that arises from latitudinally-averaging the thermal inertia and emissivity data from \citet{trumbo+2018}, but may be more sensitive to variations in surface temperature and/or radiation flux.

\section{Results \& Discussion}
\label{sec:res_disc}

Using full-disk ground-based observations from \citet{grundy+1999}, \citet{busarev+2018}, and the APO TripleSpec instrument, concurrently with laboratory spectra from \citet{mastrapa+2008} and the ISL, we calculate Europa's leading hemisphere crystallinity percentage as $\sim27$--$36\%$ (for the \citealp{grundy+1999}, \citealp{busarev+2018} 5-point moving window, and APO Triplespec data), and $\sim65$--$71\%$ (for the original \citealp{busarev+2018} data). The full-disk crystallinity percentage as approximated by thermophysical and radiation flux modeling with ICICLE is $\sim80$--$95\%$. This crystallinity discrepancy between the ground-based observations/laboratory derived values and the simulated crystallinity as derived from temperature and radiation flux modeling may be a result of several contributions. We explore a few possible sources for this discrepancy below.

\begin{figure}[!ht]
 \centering
 \includegraphics[scale=0.65]{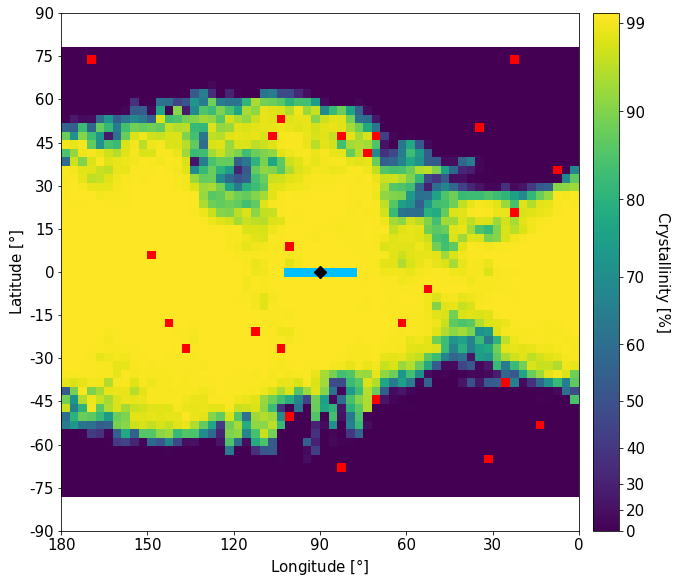}
 \caption{ICICLE-generated crystallinities of the leading hemisphere of Europa for the model run in which thermal inertias and emissivities were extended above those provided by \citet{trumbo+2018} with latitudinal averages. Overlayed are symbols indicating the three plume deposition trials; one circular region of amorphous water ice plume material with a diameter of $50$ km deposited at $0\degree$ latitude, $90\degree$ longitude (black diamond); one $80$ km $\times$ $600$ km rectangular region centered at $0\degree$ latitude, $90\degree$ longitude (blue rectangle); and $25$ randomly placed, $50$ km diameter circular plume regions (red squares).}
 \label{fig:xtal_plumes}
\end{figure}

Surface deposits of vapor-deposited plume material are likely, given the recent possible detections of plume outbursts \citep[e.g.,][]{roth+2014,sparks+2017} and water vapor above the surface \citep{paganini+2019}. In an attempt to better understand the possible contribution of vapor deposition to the full-disk crystallinity, we simulated depositing amorphous water ice plume material onto the surface to determine whether its presence could lower the crystallinity by a detectable amount. In order to maximize its effect on the full-disk crystallinity, we simulated this amorphous water ice deposit instantaneously (i.e., it has not yet thermally relaxed into the crystalline form) at the sub-observer location ($0\degree$ latitude, $90\degree$ longitude) so that it experienced the maximum spatial weight when performing a projection area-weighted average across the full leading hemisphere disk, and we assumed it coated the surface densely enough such that the full area of the deposit is $100\%$ amorphous water ice. When we emplaced a circular region of amorphous water ice plume material with a diameter of $50$ km \citep{southworth+2015}, the modeled full-disk crystallinity decreased by $0.07\%$. For an $80$ km $\times$ $600$ km rectangular region \citep[which could surround a linear crack;][]{sparks+2017}, the modeled full-disk crystallinity decreased by $0.45\%$. Finally, for $25$ randomly placed, $50$ km diameter circular plume regions, the modeled full-disk crystallinity decreased by $1.02\%$ (Figure \ref{fig:xtal_plumes}).

These changes to the full-disk crystallinity are well below the uncertainty of the modeled and observed crystallinities and would therefore be undetectable. Even if one assumes that these plumes have been erupting and depositing amorphous water ice onto the surface for the past $100$ Myr, this plume activity and its deposition of amorphous water ice would be largely erased at equatorial and mid-latitudes due to the short thermal relaxation timescale required for deposited amorphous water ice to convert into crystalline water ice \citep{fama+2010}. Deposited amorphous water ice would likely crystallize in an even shorter amount of time if the ice has substantial porosity compared to compact ice \citep{mitchell+2017}. Furthermore, assuming a large portion of the erupted amorphous material condenses near the poles, as proposed by \citep{hansen+mccord2004}, the full-disk crystallinity is unlikely to decrease significantly because much of the water ice at higher latitudes is already predominantly in the amorphous form due to lower temperatures, and regions at higher latitudes contribute minimal signal in a disk-averaged view.

Events that may have temporarily increased the surface temperature, such as meteorite impacts or cryovolcanic eruptions, would only increase the crystallinity, while there are only a few processes that would decrease the crystallinity, such as vapor-deposited plume material. Therefore, even if these physical processes were included, the modeled crystallinity would likely only increase rather than decrease.

To further elucidate possible sources for the discrepancy between the observed and modeled crystallinities, we consider the extent to which several model parameters must change in order to decrease the modeled crystallinity by a substantial, detectable amount. The modeled crystallinity decreases by $\sim9\%$ when we increase the ion flux by a factor of 10 (see Table \ref{table:model_cryst}). Therefore, an ion flux onto the leading hemisphere that greatly exceeded the ion flux onto the trailing hemisphere would be required to decrease the modeled crystallinity by a substantial, detectable amount. This is highly unlikely, as \citet{johnson+1988} demonstrate that the trailing hemisphere receives $\sim5-15$ times more ion flux than the leading hemisphere.

Another possibility that could explain the discrepancy between the observed and modeled crystallinities may originate in the simulated surface temperatures. While many of the parameters that influence the simulated surface temperatures include various uncertainties to account for deviations in thermal inertia, emissivity, and albedo (see Section \ref{sec:model}), we still consider inaccurate temperature values as a possible source for the crystallinity discrepancy. The modeled crystallinity decreases by $\sim7\%$ when we decrease the average surface temperature at each model gridpoint by $3$ K. In order to decrease the modeled crystallinity enough to begin approaching the observed crystallinities, the average surface temperature at each gridpoint location in ICICLE would need to be decreased by more than $10$ K. This is unlikely, as our daily temperature profiles (e.g., Figure \ref{fig:temp_latvstod}) are in agreement with previously published modeled temperature profiles \citep[e.g.,][]{spencer+1999,trumbo+2017}.

Several other possible sources for the discrepancy between the modeled and observed crystallinities include:

\begin{enumerate}
  \item The observed ground-based data are reflectance spectra, whereas the experimental laboratory data from \citet{mastrapa+2008} and the ISL at JPL are both transmission spectra. Differences in the methods of acquiring these types of spectra may contribute to an inaccurate calculation of the integrated band areas.
  \item Variations in water ice grain size across Europa's surface could impact the crystallinity calculation, since band depths increase with grain size \citep[e.g.,][]{clark1981,jaumann+2008}. Although we use band area ratios in an attempt to remove grain size dependency, variations in ice grain size across Europa's surface could still impact the crystallinity calculations.
  \item Spatially-distributed ion fluxes at the leading hemisphere have yet to be published, therefore our use of a constant ion flux across the surface may produce an erroneous calculation of the fraction of amorphous water ice at each location on the surface.
  \item Global water vapor distribution due to sublimation of ice and/or migration of plume material may occur at higher latitudes, where temperatures as low as 60 K occur. More amorphous water ice is likely to dominate these regions, in agreement with the laboratory-derived crystallinity values. The small projection area-weighted factor these regions provide to the disk-integrated crystallinity suggest that high latitude processes alone are unlikely to explain the crystallinity discrepancy.
\end{enumerate}


\section{Conclusions}
\label{sec:concl}

Thermal relaxation, ion particle bombardment, and geophysical processes can alter the crystallinity of Europa's surface. Comparison of observation-derived and numerically-modeled calculations of Europa's surface water ice crystallinity can provide context for the extent to which each of the above surface processing mechanisms might have an effect on Europa's surface. We derive from ground-based spectroscopic full-disk observations and laboratory water ice measurements crystallinity ranges of $\sim27$--$36\%$ for the \citet{grundy+1999}, \citet{busarev+2018} 5-point moving window, and APO TripleSpec data, and $\sim65$--$71\%$ for the original \citet{busarev+2018} data. However, we model a crystallinity of $\sim80$--$95\%$ when incorporating effects from thermal relaxation and ion particle bombardment. This discrepancy between the observed and modeled crystallinity may be a result of uncertainties in the calculations of either quantity. Although it appears unlikely that the presence of plume-deposited amorphous water ice could decrease the modeled crystallinity a substantial, detectable amount, we do not immediately rule out geophysical processes as possible contributions, as there is still much unknown about these processes on Europa. Future work will involve analyzing the water ice crystallinity at smaller scales on Europa's leading hemisphere using spectral imaging from the \emph{Galileo} Near-Infrared Mapping Spectrometer instrument, and contextualizing their surface emplacement into our understanding of Europa’s full-disk leading hemisphere crystallinity. Our study also opens up an opportunity to conduct ground-based observations of Europa on a routine basis to monitor any significant changes that may take place over time.

\section*{Acknowledgments}
\label{ack}

The authors would like to thank Samantha Trumbo for providing thermal inertia, emissivity, and albedo maps for use in our thermophysical model, Eleonora Ammannito for conducting the water ice experiments at ISL JPL, and two anonymous reviewers for helpful comments that improved the quality of this manuscript. This study was funded by NASA under Grant 80NSSC17K0408 issued through the NASA Education Minority University Research Education Project (MUREP) as a NASA Harriett G. Jenkins Graduate Fellowship through the Aeronautics Scholarship \& Advanced STEM Training and Research (AS\&ASTAR) Fellowships. A part of this work was carried out at the Jet Propulsion Laboratory, California Institute of Technology, under a contract with the National Aeronautics and Space Administration (NASA). Funds were provided through NASA Solar System Workings Program.



\bibliographystyle{elsarticle-harv}
\section*{References}
\bibliography{europa_bibl}



\end{document}